\documentclass{article}

\usepackage{arxiv}

\usepackage[utf8]{inputenc} 
\usepackage[T1]{fontenc}    
\usepackage{hyperref}       
\usepackage{url}            
\usepackage{booktabs}       
\usepackage{amsfonts}       
\usepackage{nicefrac}       
\usepackage{microtype}      
\usepackage{lipsum}

\usepackage{morefloats}
\usepackage{color}
\usepackage[table]{xcolor}
\usepackage{multirow}
\usepackage{latexsym}
\usepackage{amsmath,amssymb,amsthm,graphicx}
\usepackage{dsfont}
\usepackage{enumitem}
\usepackage{hyperref}
\usepackage{arydshln}
\usepackage{lscape}
\usepackage{adjustbox}
\newtheoremstyle{thm}
{9pt}
{9pt}
{\itshape}
{}
{\bfseries}
{.}
{ }
{}
\theoremstyle{thm}
\DeclareMathAlphabet{\mathscr}{OT1}{pzc}{m}{it}

\newtheoremstyle{def}
{9pt}
{9pt}
{}
{}
{\bfseries}
{.}
{ }
{}
\theoremstyle{def}

\renewcommand{\footnoterule}{%
	\kern -3.5pt
	\hrule width \textwidth height 1pt
	\kern 3.5pt
}

\makeatletter
\def\blfootnote{\xdef\@thefnmark{}\@footnotetext}
\makeatother

\title{A statistical technique for cleaning option price data}

\author{
I.J.H. Visagie\\
Pure and Applied Analytics,\\ North-West University,\\ South Africa.\\
\href{mailto:jaco.visagie@nwu.ac.za}{jaco.visagie@nwu.ac.za}\\
}

\begin{document}

\date{\today}
\maketitle

\begin{abstract}
    Recorded option pricing datasets are not always freely available. Additionally, these datasets often contain numerous prices which are either higher or lower than can reasonably be expected. Various reasons for these unexpected observations are possible, including human error in the recording of the details associated with the option in question. In order for the analyses performed on these datasets to be reliable, it is necessary to identify and remove these options from the dataset. In this paper, we list three distinct problems often found in recorded option price datasets alongside means of addressing these. The methods used are justified using sound statistical reasoning and remove option prices violating the standard assumption of no arbitrage. An attractive aspect of the proposed technique is that no option pricing model-based assumptions are used. Although the discussion is restricted to European options, the procedure is easily modified for use with exotic options as well. As a final contribution, the paper contains a link to six option pricing datasets which have already been cleaned using the proposed methods and can be freely used by researchers.\footnote{I thank Prof F Lombard for valuable contributions to this work. Unfortunately Prof Lombard passed away before the paper was completed, any remaining errors are my own.}
\end{abstract}


\section{Introduction and motivation}
\label{Intro}


Option pricing is both of great practical and theoretical interest. The practical importance is easily seen when considering the vast size of the global market in options, while the theoretical importance is informed by the numerous papers relating to option pricing, for more details, see \cite{CT2004} and \cite{Sch2003}. When new option pricing models or calibration techniques are proposed, or existing techniques are compared, observed option prices are typically required in order to demonstrate the use of the proposed models or techniques, see \cite{SST2004}.

Observed option price data can often be problematic as the recorded datasets are not always freely available and frequently contain errors of various kinds. For instance, even when these datasets can be obtained, some of the option prices in the captured data may appear to be either higher or lower than can reasonably be expected. We believe that this could be the result of human error (including typing mistakes) or some other shortcoming of the method used to capture the data. We believe that, if these unexpected option prices are allowed to remain in the datasets, the empirical results obtained will be negatively impacted. In order to assure that the analyses performed on these datasets are accurate, it is important to identify such problems and remove the affected options. One method of circumventing errors of this kind is to define a so-called ``theoretical'' market, see \cite{Vis2018b}. However, in most cases, it would be preferable to use observed data.


A vast literature on data cleaning is available including numerous textbooks as well as scholarly papers, see, for example, \cite{DJ2003}, \cite{GS2013}, \cite{KL2022} and the references therein. Many fields of inquiry even boast tailored approaches specifically designed for use in the discipline in question. As a specific example found in manufacturing, see \cite{MMK2024}. Surprisingly, we were unable to find any papers specifically relating to the cleaning of option price data as observed at a given point in time. However, it should be noted that several papers on related subjects are available, including algorithms for the detection of outliers in ultra high-frequency financial data, see \cite{VaG2010}.

The current paper proposes a data cleaning technique specific to option pricing datasets, as recorded on a specific date, which is both logically consistent and statistically sound. By logically consistent, we mean that application of the proposed method removes any arbitrage opportunities from the market. The statistical aspects of the technique is explained below. We describe three steps used to remove the relevant options from the datasets. The first step is to remove option prices that lead to arbitrage opportunities. The second is to identify and remove outlying option prices and the final step is to remove duplicated options.


An attractive aspect of the proposed technique is that it is non-parametric in the sense that is free of any option pricing model-based assumptions. That is, the approach advocated for is not restricted to use with a specific option pricing model, but can be applied regardless of the choice of model. Regarding the type of option under consideration, we restrict our attention to European call and put options below. However, the proposed techniques are general and can, in principle, be used in conjunction with exotic options as well.

As was mentioned earlier, it is often difficult to obtain suitable datasets which can be used in conjunction with research relating to option pricing. In an attempt to be of use to other researchers in the field, we provide six datasets which can be used for this purpose. These datasets can be downloaded from \url{https://github.com/JacoVisagieNWU/Option-pricing-data}. The details of these datasets are discussed in the following section. While these option price datasets are not recent, they can be used without any limiting issues related to confidentiality (as is often the case for more recent datasets). Furthermore, these datasets relate to prices recorded in May of 2012, which is towards the end of the financial crisis starting in 2010. As a result, in the opinion of the author, these datasets may be of considerable interest to researchers as they relate to a time in which financial markets were considered to be fairly volatile.

The remainder of the paper is structured as follows. The next section provides the details of six option pricing datasets. Section 3 constitutes the methodological contribution, it lists three distinct problems often found in recorded option price datasets alongside means of addressing these. Section 4 provides some final results and conclusions.


\section{Option pricing datasets considered}

The options considered are obtained from the American financial market, which is the largest and most liquid financial market in the world. As a result, it is hard for a single market participant to manipulate option prices. The data used consist of European call and put option prices on the S\&P 500 index, the PowerShares index and Google shares. The market data discussed is as at close of business on 11 May 2012. The six datasets considered were obtained from \url{http://finance.yahoo.com}.

The values of a range of economic variables are required in order to fit an option pricing model. These include the risk-free interest rate and the rate at which the stock in question yields dividends.
Since all of these datasets are obtained at the same time and in the same market, the same risk free interest rate will be used for each dataset.
We use the discount rate of Treasury bills in the secondary market as the risk free interest rate. For each of the datasets below the average of the times to maturity is roughly six months, so we use the discount rate on the six month treasury bill as the risk free interest rate. On 11 May 2012 this rate was 0.15\% per annum. Two of the three assets underlying the option prices discussed above are indexes which do not pay dividends. Google is one of the few traded companies that do not pay out any dividends. As a result, the dividend rate is 0 for all underlying assets considered.

On 11 May 2012 the S\&P 500 index closed at \$1353.39. We obtained the prices of 576 call options and 779 put options on this index from \cite{SP500}. The PowerShares index is an index weighted by market capatilisation which consists of the largest non-financial companies the world over. On this day the PowerShares index closed at \$64.18. The recorded options on this index include 413 call options and 480 put options, see \cite{PowerShares}. On the same date the share price of Google Inc. closed at \$605.23. We obtained the prices of 545 call options and 532 put options on this stock from \cite{Google}.

When fitting an option pricing model, one is usually required to estimate the probability measure underlying the market, see \cite{VG2019}. This estimate is typically based on the log-returns of the price of the underlying asset. In order to calculate these log-returns, the historical prices of the stock and the indexes underlying the options were obtained from \url{http://finance.yahoo.com}. For a discussion on parametric means of this type of estimation based on observed log-returns, see \cite{Vis2018}. When deciding on the duration of the historical period in which to consider the observed log-returns, a balance must be struck between the desire for a large dataset and the relevance of the observed prices. The historical period considered is used to gain information on the (constantly changing) current market conditions. If this period extends too far into the past it is possible that the market conditions have changed considerably within this historical period. In order to furnish researchers with the information required in order to fit option pricing models, we provide the log-returns of the stock and indexes in question for a period of one year prior to the date on which the option prices were recorded.

The six datasets discussed above can be downloaded in the form of Microsoft Excel files from \url{https://github.com/JacoVisagieNWU/Option-pricing-data}. These files are clearly labelled and, for each option, contain the option price, the strike price and the time to maturity. As was mentioned above, the calculated log-returns are also included. Note that the options recorded in these datasets and those remaining after the data cleaning process discussed below.


\section{Techniques for the identification and treatment of errors\label{The data}}


The typically information recorded alongside the price of a given option include the option type, the time to maturity, the strike price and the open interest associated with the option. The open interest of an option is the number of open positions currently held by investors in that option, this is a measure of how actively the option is traded. Based on these recorder values, we aim to identify problematic option prices.

Below we describe three distinct problems encountered when analysing observed data as well as the methods proposed to identify the erroneously recorded values. We illustrate the proposed techniques using practical examples throughout. All of the calculations are performed using Matlab 2022b.

\subsection{Option prices leading to arbitrage opportunities}


Upper and lower bounds for arbitrage free European call and put option prices are provided in \cite{Hul2009}. These bounds are not model specific and the arbitrage free intervals provided below are the largest intervals possible under any option pricing model. The bounds for a European call option price, $\pi ^{c}$, are
\begin{equation}
    \textrm{max}\left( S_{0}-Ke^{-rT},0\right)\leq \pi ^{c}\leq S_{0}, \label{bounds for call price}
\end{equation}
and the bounds for a European put option price, $\pi ^{p}$, are
\begin{equation}
    \textrm{max}\left( Ke^{-rT}-S_{0},0\right) \leq \pi ^{p}\leq Ke^{-rT}. \label{bounds for put price}
\end{equation}

If a European call option price falls outside the interval given in (\ref{bounds for call price}), then this option constitutes an immediate arbitrage opportunity. Similarly a European put option price not contained in the interval (\ref{bounds for put price}) provides an immediate arbitrage opportunity. Since the options considered are actively traded, it is reasonable to assume that these arbitrage opportunities would be exploited immediately. If, for example, an option price falls below the lower bound specified above, then traders would buy vast numbers of these options in order to realise risk free profits. As a result, the demand for this option would exceed the supply and normal market forces would increase the price of the option until the arbitrage opportunity is removed. Similarly, if an option price was to exceed the upper bound, then investors would sell this option in large quantities and the supply would exceed the demand, this would lower the price and remove the arbitrage opportunity.

In order to ensure that the option prices used are realistic, we remove all call options from the dataset that do not satisfy (\ref{bounds for call price}). Similarly we remove all put options that fail to satisfy (\ref{bounds for put price}).

\subsection{Outlying option prices}

After removing the option prices not contained in realistic intervals we turn our attention to possible outliers remaining in the data. For each dataset the following method is used to identify outliers and to remove these option prices from the datasets. For each dataset the options are split into several smaller datasets according to their time to maturity; i.e. we group all options with equal times to maturity together. We fit a second degree polynomial (using linear regression) to each of these groups of option prices; the strike price of the option is used as the predictor variable and the option price is used as the response. A second degree polynomial is used to compensate for the volatility smile, see \cite{FPS2001}. 
Figure 1 shows the prices of the call options on the Google shares with a time to maturity of 160 days with the fitted polynomial superimposed.

\begin{figure}[!htbp!] \label{fig1}
    \begin{center}
        \includegraphics[width=0.7\textwidth]{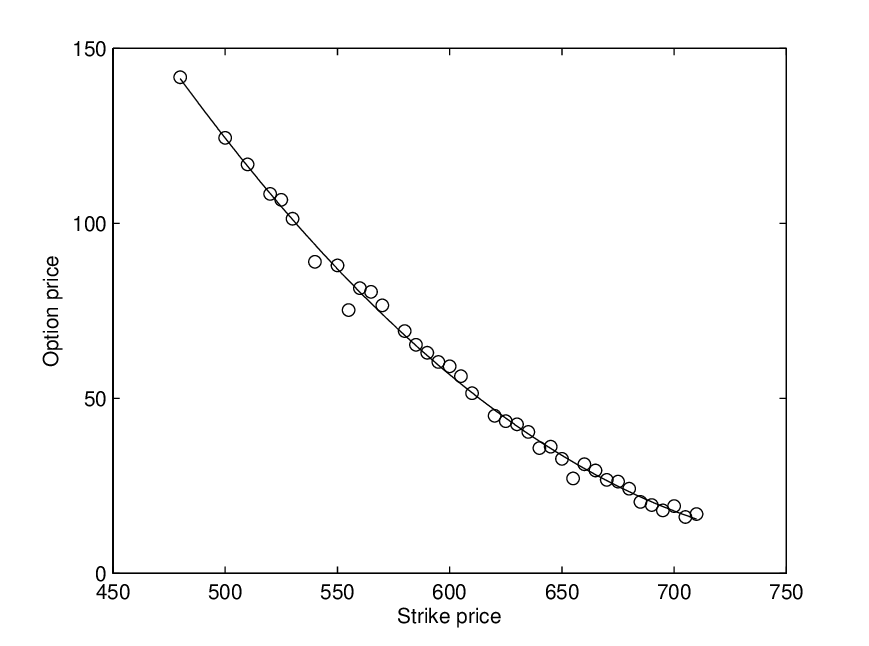}
    \end{center}
    \caption{Google call option prices with the fitted second degree polynomial.}
\end{figure}


We calculate the residuals associated with the regression. Outliers in the option prices are identified by identifying outliers in the residuals using a confidence interval, this interval is derived below.

Consider the residuals from a given group of options with a common time to maturity. Denote the number of residuals in this group by $n$ and the $j^{\text{th}}$ residual by $r_{j}$; $j=1,...,n$. Assume that the residuals are independently and identically distributed normal random variables with a mean of $0$ and standard deviation of $\sigma $. Let $\alpha \in \left(0,1\right) $ be the probability that there is at least one residual in the group not contained in some interval $\left[ -c,c\right] $ with $c>0$. Let $A$ be the event that the group considered contains no such residuals;
\begin{equation}
    P\left( A\right) =1-\alpha .  \label{PA1}
\end{equation}
From independence
\begin{eqnarray}
    P\left( A\right) &=&\prod_{j=1}^{n}P\left( -c\leq r_{j}\leq c\right)  \notag \\
    &=&\left[ P\left( -c\leq r_{1}\leq c\right) \right] ^{n}  \notag \\
    &=&\left[ 2\Phi \left( \frac{c}{\sigma }\right) -1\right] ^{n},  \label{PA2}
\end{eqnarray}
where $\Phi$ denotes the standard normal distribution function. Equating (\ref{PA1}) and (\ref{PA2}) we obtain

\begin{equation*}
    \alpha =1-\left[ 2\Phi \left( \frac{c}{\sigma }\right) -1\right] ^{n},
\end{equation*}%
which can be solved for $c$:%
\begin{equation}
    c=\sigma \Phi ^{-1}\left( \frac{1}{2}+\frac{1}{2}\left( 1-\alpha \right)^{1/n}\right) .  \label{c formula}
\end{equation}

In the argument used above we assume that $\sigma $ is known. However, this is not the case. In order to implement this procedure we estimate $\sigma $ using the standard deviation of the realised residuals. This value is used in equation (\ref{c formula}) in the place of $\sigma $ to obtain an estimate for $c$ (denoted $\hat{c}$). Any residual not contained in the interval $\left[ -\hat{c},\hat{c}\right] $ is deemed an outlier and the corresponding option is removed from the dataset.

We implement the method with $\alpha =0.01$. Figure 2 shows the residuals associated with the call options on the Google shares with a time to maturity of 160 days. The horizontal lines represent $\hat{c}$ and $-\hat{c}$ respectively. In this example a single residual lies outside the interval $\left[ -\hat{c},\hat{c}\right] $. This residual is deemed an outlier and the corresponding option price is removed from the dataset.

\begin{figure}[!htbp!] \label{fig2}
    \begin{center}
        \includegraphics[width=0.7\textwidth]{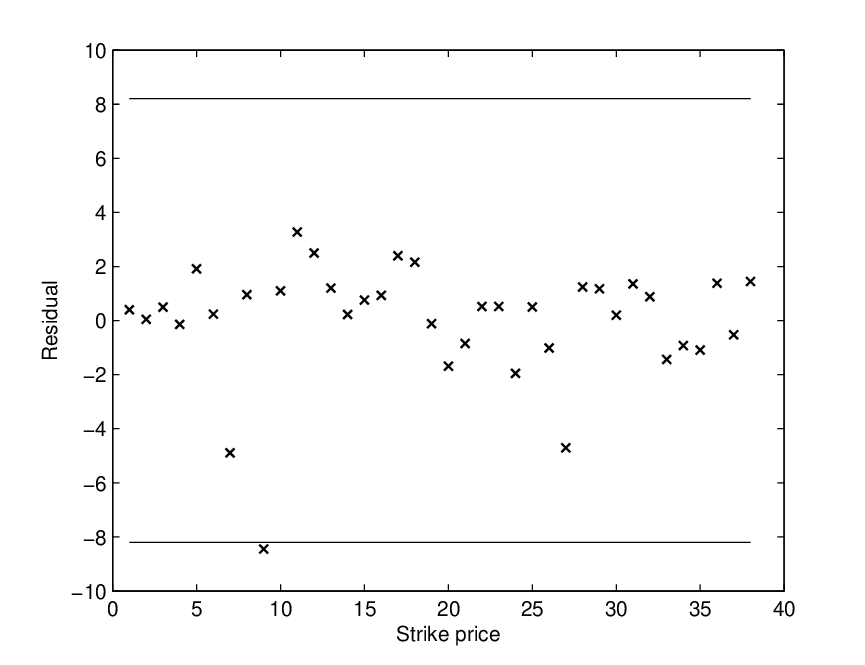}
    \end{center}
    \caption{Residuals together with an estimated 99\% confidence interval.}
\end{figure}


\subsection{Duplicated options}

The prices of some of the recorded options are not unique, i.e. there are multiple options with equal strike prices and times to maturity but different prices. It is not possible for an option to have more than one price, since this would lead to an instant arbitrage opportunity. If an option had two different prices traders would buy the option at the cheaper price and sell it at the more expensive to realise an immediate risk free profit. As a result, the demand for the cheaper option and the supply of the more expensive option would increase. This in turn would force the price of the cheaper option to rise, while the price of the more expensive option would decrease. This process would continue until the option prices were equal.

A European call (put) option, with given time to maturity, is a decreasing (increasing) function of the strike price. As before, if these conditions do not hold, it will lead to an arbitrage opportunity. It is therefore reasonable to enforce this condition in the data used. All options with multiple prices are identified. The prices of these options are compared to the prices of the options with strike prices immediately smaller and larger than their own strike price (where all options considered have the same time to maturity). In the case of a call (put) option with multiple prices, the option is removed from the dataset if its price is larger (smaller) than that of the option with strike price immediately larger than its own, or if its price is smaller (larger) than that of the option with strike price immediately smaller than its own.

The procedure described above does not remove all of the duplicates in the data. To remove the remaining duplicates we consider the recorded open interest of these options.
We reason that the more actively an option is traded, the less likely it is that the recorded price of the option will be incorrect. As a result, for each unique strike price and time to maturity, all of the duplicated options are removed save the one with the largest open interest.

\section{Results and conclusions}

After applying the methods described above the resulting datasets are as follows. The S\&P 500 dataset consists of 430 call options and 605 put options. The original sample sizes, before the data cleaning process, were 576 call options and 779 put options. This means that 146 call options and 129 put options were removed during the cleaning process, which corresponds to 25\% and 17\% of the original datasets, respectively.

Turning our attention to the PowerShares dataset, we see that the data cleaning process reduces the number of call options from 413 to 293, which means that 29\% of the options were removed. Of the original 480 put options 199 are removed, resulting in a dataset containing 281 option prices. In this case 41\% of the option prices are removed by cleaning.

For the Google Inc. dataset, the number of call options is reduced from 545 to 519, constituting a 5\% reduction in the number of options in the dataset. In the case of put options, 87 of the 532 options are removed, resulting in a dataset consisting of 445 options.
The number of options used is reduced by 16\% in this instance.

For the six datasets considered, the large number of options removed in each case attests to the need for a data cleaning technique specific to this field. In the most extreme case (that of the PowerShares put options), it was necessary to remove 41\% of the recorded options in order to ensure that the prices are realistic.

\bibliographystyle{plain}
\bibliography{lit-AC}

\end{document}